\documentclass{aa}

\usepackage{lineno}
\usepackage{hyperref}
\usepackage{siunitx}
\usepackage{booktabs}
\usepackage{tabularx}
\usepackage{array}
\usepackage{graphicx}
\usepackage{subfig}
\usepackage{afterpage}
\usepackage{bm}
\usepackage{hyperref}
 \usepackage{orcidlink}

\usepackage{caption}  
\usepackage{sidecap}  
\usepackage{lipsum} 
\usepackage[font=small]{caption} 

\usepackage{txfonts}
\usepackage{hyperref}

\begin{document} 

   \title{Physical properties of newly active asteroid 2010 LH$_{15}$}

   \author{Bin Liu\inst{1, 2}\orcidlink{0009-0004-3405-9235}
          \and Cunhui Li\inst{3}\orcidlink{0000-0003-2689-9387}
          \and Zhongcheng Mu\inst{4}\orcidlink{0000-0002-8868-0447}
          \and
          Xiaodong Liu\inst{1, 2}\orcidlink{0000-0001-9329-0315}
          }
   \institute{School of Aeronautics and Astronautics, Shenzhen Campus of Sun Yat-sen University, Shenzhen, Guangdong 518107, China    \and
             Shenzhen Key Laboratory of Intelligent Microsatellite Constellation, Shenzhen Campus of Sun Yat-sen University, Shenzhen, Guangdong 518107, China   \and
             Science and Technology on Vacuum Technology and Physics Laboratory, Lanzhou Institute of Physics, Lanzhou, 730000, China   \and
             School of Aeronautics and Astronautics, Shanghai Jiao Tong University, Shanghai, 200240, China\\
             \email{liuxd36@mail.sysu.edu.cn}
             }

   \date{}

 
  \abstract
{Main-belt asteroid 2010 LH$_{15}$ has been classified as an active asteroid, based on the recent discovery of dust activity from the archival images observed in 2010 and 2019. In this study, we perform measurements and dynamical modeling of the dust tail of the active asteroid 2010 LH$_{15}$ using ZTF archival data from July 26 to August 31, 2019, with the derived physical properties from these relatively independent methods being compatible. The photometric results show that the radius of the nucleus is $1.11\pm0.02$ km with assumed geometric albedo of $p_r = 0.05$, and the color index of the nucleus is relatively close to that of the ejecta around the nucleus, with a value of $H_g - H_r = 0.44\pm0.07$. The effective scattering cross-section increases at an average rate of $0.28\pm0.02$ km$^2$ day$^{-1}$ throughout the observation period, indicating that the activity of LH$_{15}$ is likely driven by mechanisms capable of causing a sustained process like sublimation. Further dust dynamics modeling indicates that the dust activity initiates as early as about 26 June 2019, with the ejected dust particles having a radius ranging from 0.03 mm to 3 mm. The dependence of the terminal velocity on dust size is consistent with a sublimation-driven mechanism. If the orbit of LH$_{15}$ is stable, its sublimation origin will extend the inner boundary of the water-ice-bearing region in the main asteroid belt inward by approximately 0.1 AU.}
   
   
   


   \maketitle
%
\section{Introduction}
Active asteroids, a newly recognized class of minor celestial bodies in the Solar System, are dynamically classified as asteroids but exhibit cometary appearance under certain conditions \citep{jewitt2012active}. As of 31 December 2023, a total of 45 active asteroids have been discovered, with the majority orbiting in the main asteroid belt. The activity of active asteroids may be driven by non-sublimation mechanisms, such as impact (e.g.~Dimorphos \citep{cheng2023momentum}) and rotational instability (e.g.~(6478) Gault \citep{kleyna2019sporadic}), or by sublimation of water ice (e.g.~238P \citep{hsieh2011main,kelley2023spectroscopic}). In particular, the recurrent activity of active asteroids are suspected to be driven by the sublimation of water ice, offering evidence supporting the hypothesis of the presence of water ice in the main asteroid belt, and thereby providing a perspective for assessing the role of main-belt asteroids in the delivery of water to the early Earth \citep{maclennan2012nucleus}. Consequently, the research of active asteroids suspected to be driven by sublimation of water ice can help to provide insights into the distribution of volatile water ice in the Solar system and offer clues about the origins of these volatiles and their potential delivery to early Earth. In addition, as the sublimation of water ice proceeds, the material from the active asteroids is pushed by the resulting volatiles out of the surface, leading to the formation of the dust structures. The research on these dust structures helps deepen scientific understanding of the formation and evolution of active asteroids \citep{ivanova2023long}.

The asteroid 2010 LH$_{15}$, hereafter "LH$_{15}$", is a middle main-belt asteroid. As of 13 September 2023, the orbital elements of LH$_{15}$ obtained from JPL\footnote{\url{https://ssd.jpl.nasa.gov/tools/sbdb_query.html}} are the semi-major axis of $a=2.743$\,AU, eccentricity of $e=0.355$, and inclination of $i=10.907\degr$, resulting in a Jupiter Tisserand parameter of $T_\mathrm{J}=3.23$. The asteroid LH$_{15}$ was discovered to exhibit a cometary appearance in one DECam image and other archival images obtained during two epochs of 27 September-10 October 2010 and 10 August-3 November of 2019, which correspond to the periods when LH$_{15}$ was near its perihelion \citep{chandler2023new}. In the archival images \citep{chandler2023new}, it is observed that LH$_{15}$ exhibits a tail during its active periods, extending nearly aligned with the anti-solar direction, which is a characteristic feature of a dust tail, indicating the occurrence of the dust activity of LH$_{15}$. Consequently, given the asteroidal orbit and dust activity of LH$_{15}$, it is reasonable to classify it as an active asteroid. LH$_{15}$ is expected to reach its perihelion passage on 26 March 2024 and the dust activity might occur before it reaches its perihelion. Research on the physical properties of LH$_{15}$ can help to assess the mechanisms of the possible forthcoming dust activity and deepen scientific understanding of this object as a member of the active asteroid class.

In this study, the archival observational data provided by ZTF (Zwicky Transient Facility; \citealt{bellm2018zwicky}) are analyzed, and the observed morphology and brightness of the dust tail are reproduced using a dust dynamics model to characterize the dust activity of LH$_{15}$. This paper is organized as follows. The archival images are presented in Section \ref{Data}. The results of the measurements and dynamical modeling of the dust tail are shown in Section \ref{Results}. The activity mechanism is discussed in Section \ref{discussion}. Finally, the conclusions are presented in Section \ref{CONCLUSIONS}.

\section{Data}
\label{Data}
The source of observational data of LH$_{15}$ analyzed in this study, ZTF, is an astronomical sky survey located at the Palomar Observatory. The purpose of ZTF is to search for objects like supernovae, asteroids, and comets, using a 48-inch Schmidt-type telescope equipped with a 600-megapixel CCD camera, giving a field of view of 47 deg$^2$ and an image scale of $1\arcsec$ pixel$^{-1}$ for each exposure \citep{graham2019zwicky}. 

\begin{figure*}
\centering
    \includegraphics[width=2\columnwidth]{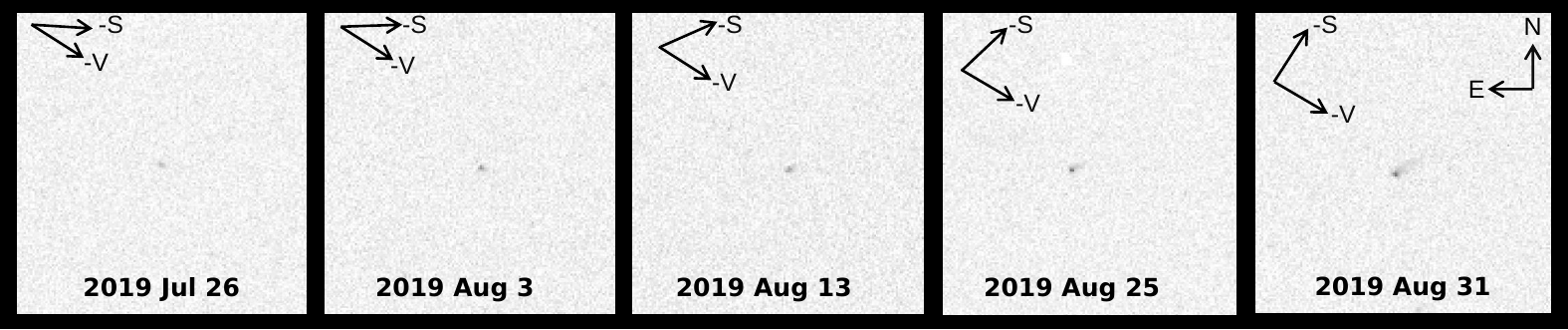}
    \caption{Composite images of LH$_{15}$. The observation epoch is marked in the upper right corner of each panel. The projected Sun-LH$_{15}$ radial direction (-S) and the anti-velocity direction of LH$_{15}$’s projected motion (-V) are indicated as red arrows. The FOV of each image is $2.5\arcmin\times2.5\arcmin$.} 
    \label{ObsImag}
\end{figure*}

ZTF has made over 100 observations of LH$_{15}$ between 2018 and 2019 and captured signs of its dust activity on 26 July 2019. Figure \ref{ObsImag} presents the dust morphology of LH$_{15}$ on 26 July, 3 August, 13 August, 25 August, and 31 August of 2019, during which LH${15}$ exhibited dust activity. The images are taken through the $r$ filter with the exposures of 30 seconds. The images of LH$_{15}$ taken on five different dates shown in Figure \ref{ObsImag} are first cropped to $4\arcmin\times4\arcmin$, and the field stars and rays are masked over the whole cropped images. The data reduction is carried out using the SAOImage DS9 \citep{joye2003new} and Astroart software \citep{nicolini2003astroart}. The geometries of the observational data on the five epochs are listed in Table~\ref{geometry}.

\begin{table*}
    \centering
    \renewcommand{\arraystretch}{1.2}
    \caption{Observation geometries of LH$_{15}$}
    \label{geometry}
    \setlength{\tabcolsep}{10pt}
    \begin{tabular}{lccccccccc}
		\toprule
Date & MJD$^a$ & $R^b$ & $\Delta^{c}$ & $\alpha^d$ & PsAng$^e$ & PsAMV$^f$ & $\nu^{g}$ & $\delta^{h}$ & Telescope\\
      (UT)  &(days) & (AU) & (AU) & (deg) & (deg) & (deg) & (deg) & (deg) & \\
		\hline
26 Jul 2019 & 58690 & 1.80 & 0.95 & 24.3 & 94.3 & 122.5 & 338.5 & 10.8 & ZTF\\
3 Aug 2019 & 58698 & 1.79 & 0.90 & 22.1 & 88.2 & 122.6 & 342.3 & 11.9 & ZTF\\
13 Aug 2019 & 58708 & 1.78 & 0.85 & 18.8 & 77.5 & 122.3 & 347.0 & 12.9 & ZTF\\
25 Aug 2019 & 58720 & 1.77 & 0.81 & 15.1 & 57.4 & 121.0 & 352.8 & 13.4 & ZTF\\
31 Aug 2019 & 58726 & 1.77 & 0.80 & 13.7 & 43.4 & 120.1 & 355.8 & 13.3 & ZTF\\
		\hline
		\multicolumn{10}{l}{$^a$ Modified Julian Date.}\\
		\multicolumn{10}{l}{$^b$ Distance between LH$_{15}$ and the Sun.}\\
		\multicolumn{10}{l}{$^c$ Distance between LH$_{15}$ and the Earth.}\\
		\multicolumn{10}{l}{$^d$ Phase angle, Sun-LH$_{15}$-Earth.}\\
		\multicolumn{10}{l}{$^e$ Angle between the Sun-LH$_{15}$ direction and the north direction (measured from the North to the East).}\\
		\multicolumn{10}{l}{$^f$ \parbox{0.9\textwidth}{Angle between the negative projected heliocentric velocity vector and the north direction (measured from the North to the East).}}\\
		\multicolumn{10}{l}{$^g$ True anomaly of LH$_{15}$.}\\
		\multicolumn{10}{l}{$^h$ Angle of observer from the orbital plane of LH$_{15}$.}\\
	 \end{tabular}
    \end{table*}

\section{Analysis}
\label{Results}
\subsection{Morphology}
\label{morphology}

In each panel of Figure \ref{ObsImag}, the upper left corner specifies the direction opposite to the motion of LH$_{15}$ and the direction away from the Sun, while the upper right corner specifies the date of the observation. The red arrow labeled with 'N' points towards true North, and another arrow labeled with 'E' points towards true East. The observation images reveal that LH$_{15}$ gradually develops comet-like features, including a coma and a tail. The coma of LH$_{15}$ is relatively compact, with the section oriented towards the Sun comprising dust particles ejected toward the solar direction. The tail in different images always lies between the direction opposite to the motion of LH$_{15}$ and the direction away from the Sun, which serves as a distinctive characteristic of a dust tail. The length of LH$_{15}$'s dust tail is determined by identifying where the surface brightness of the tail reaches the level of the background noise, which is applied consistently to all images. For the five epochs, the tail lengths are measured to be $2.4\pm0.3$ arcseconds, $3.5\pm0.4$ arcseconds, $7.3\pm0.9$ arcseconds, $10.1\pm1.2$, and $18.5\pm1.7$ arcseconds.


A preliminary estimation of dust particle ejection velocity is derived by considering the motion of ejected particles within the sunward section of the dust coma. These particles, once ejected from the hemisphere facing the sun, are decelerated by radiation pressure until they reach the edge of the coma. Thus, the estimation of dust particle ejection velocity is inferred from the measurements of the sunward extension of the dust coma. Following Equation (4) from \citet{jewitt1987surface}, the ejection velocity of the dust particle $v_{\text{ej}}$ is estimated as
\begin{equation}
v_{\mathrm{ej}}=\sqrt{2 \beta g_{\text{sun}} l_{\text {coma }} },
	\label{vej}
\end{equation}
where $l_{\text{coma}}$ is the measurement of the sunward dust coma, i.e.~the turnaround distance of the dust particles in the sunward direction. The variable $g_{\text{sun}} = g {(1)}\sin \alpha/R^2$ is the local gravitational acceleration to the Sun, where $g{(1)}=0.006$ m/s$^2$ is the gravitational acceleration at one astronomical unit from the Sun, $\alpha$ is the phase angle, and $R$ is the heliocentric distance. The dimensionless coefficient $\beta$ represents the ratio of radiation pressure to gravitational force acting on dust particles and is inversely proportional to the grain radius. Given that the structure of the coma is most clearly visible in the observation on 31 August 2019, the ejection velocity is estimated from this observation. In this image, the sunward extent of the coma is measured to be $(5.2 \pm 0.7) \times 10^3$ km, with the uncertainty mainly due to the seeing conditions. The phase angle and heliocentric distance of the asteroid on 31 August 2019 are shown in Table \ref{geometry}. These parameters are then substituted into Equation \ref{vej}, yielding the relationship between the ejection velocity $v_{\mathrm{ej}}$ and $\beta$ as $v_{\mathrm{ej}}\sim(68.4\pm4.6)\,\beta^{1/2}$ m s$^{-1}$.

\subsection{Photometry}
\label{photometry}

\begin{figure}
\centering
    \includegraphics[width=\columnwidth]{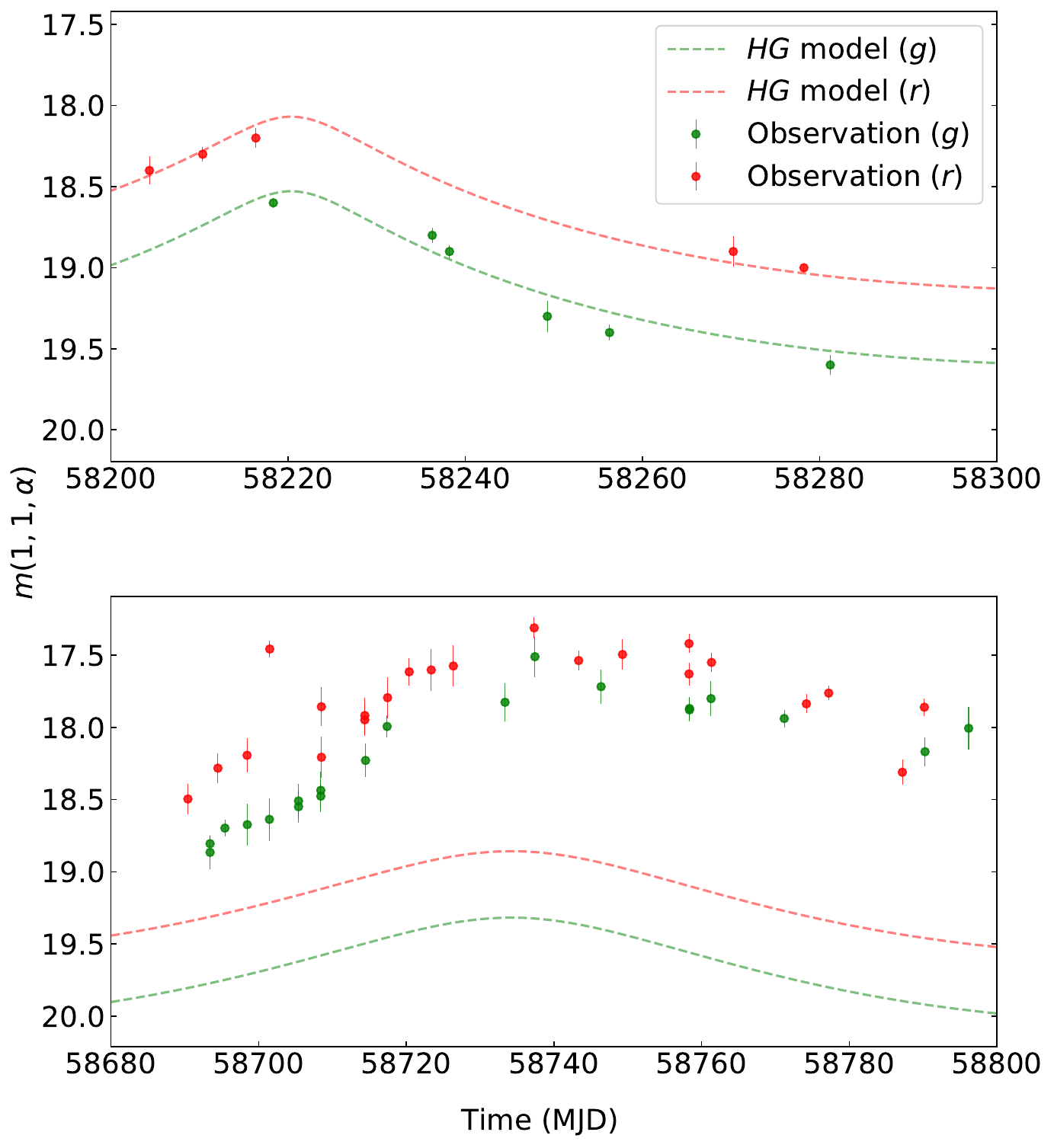}
    \caption{Reduced magnitudes of LH$_{15}$ derived from photometry (circles) and HG function (dashed line) in 2018 (upper panel) and 2019 (lower panel).}
    \label{HGmodel}
\end{figure}

The observational data of LH$_{15}$ obtained from ZTF during the years 2018 and 2019 are first analyzed within a 5\arcsec radius circular aperture, and the seeing of the images varies between 1\arcsec and 2.5\arcsec. The measured result are presented in Figure (\ref{HGmodel}). The observations acquired in 2018 (inactive periods) are selected to construct the HG model for the nucleus of LH$_{15}$. Assuming $G$ to be 0.15, $H_r$ and $H_g$ in the model are determined as $17.01\pm0.04$ and $17.45\pm0.06$, respectively. The calculated value of $H_g - H_r = 0.44\pm0.07$ suggests a higher probability that LH$_{15}$ is a C-type asteroid ($H_g - H_r \approx 0.5$) \citep{ye2019multiple}. 

Following Equation (4) from \citet{hsieh2009albedos}, the nucleus radius $r_e$ is estimated using the resulting $r$-band absolute magnitude of $H_r = 17.01\pm0.04$ by
\begin{equation}
r_{\rm e}^2=\frac{2.24 \times 10^{16}}{p_{r}}\times10^{0.4 (m_{\odot, r} - H_r)},
	\label{radius}
\end{equation}
where $r_e$ is measured in km, $m_{\odot, r} = -26.9$ is the solar $r$-band magnitude \citep{willmer2018absolute}, and the value of geometric albedo $p_r$ is set at 0.05 \citep{hsieh2009albedos}. The nucleus radius is estimated as about $r_e=1.11\pm0.02$ km.

The light curve in Figure \ref{HGmodel} indicates that the brightness of LH$_{15}$ exceeds the model's predicted brightness by approximately 1 magnitude during the initial observation in 2019 and LH$_{15}$ exhibits a slightly extended appearance. Subsequently, the brightness of LH$_{15}$ increase and culminate in a peak on September 10. This peak represents an enhancement of 1.5 magnitudes relative to the predicted values derived from the model. The photometric results during the active phase of LH$_{15}$ show the difference between $H_r$ and $H_g$ is approximately $0.50\pm0.11$. Therefore, it is assumed in the subsequent sections of this study that the phase function, albedo, and density of LH$_{15}$ exhibit characteristics of C-type asteroids.
To ascertain the physical properties of dust during the activity, it is necessary to conduct a more detailed photometric analysis of the observational data. The observation images of LH$_{15}$ obtained on 26 July, 3 August, 13 August, 25 August, and 31 August of 2019 (active periods) used for this more detailed analysis originate from the ZTF database. A set of five fixed-radius circular apertures with radii of 500, 1000, 2000, 4000, 8000 km are utilized to obtain the photometric measurements. The sky background is determined in an annular aperture with inner and outer radii of 30 arcseconds and 60 arcseconds, respectively. The resulting apparent magnitudes $m_r(R, \Delta, \alpha)$ are then converted to absolute magnitudes $H$ by

\begin{equation}
H=m_r(R, \Delta, \alpha)-5 \log _{10}\left(R \Delta\right)+2.5\log_{10}\Phi(\alpha),
	\label{magnitude}
\end{equation}
where $R$, $\Delta$, and $\alpha$ refer to Table \ref{geometry}. The phase function $\Phi(\alpha)$ employed here is consistent with the one in the HG model, with a $G$ parameter of 0.15, which is suitable for C-type asteroids. The relationship between the effective scattering cross-section, $C_\text{e}$ [km$^2$], and the absolute magnitude is derived from Equation \ref{radius} as:
\begin{equation}
C_{\rm e}=\frac{1.2 \times 10^{6}}{p_{r}} 10^{-0.4 H}.
	\label{cross-secction}
\end{equation}

Here, $p_r$ represents the geometric albedo. Assuming the albedo of the dust is consistent with that of the nucleus, the value of $p_r$ is set at 0.05, which is commonly assumed for the albedo of the active asteroids \citep{hsieh2009albedos}. For each selected epoch and aperture size, Table \ref{PhotoTable} presents computed values of the apparent magnitudes, the absolute magnitudes, and the effective scattering cross-sections. The uncertainties presented in Table \ref{PhotoTable} are dominated by the photon noise of the object. It should be noted that the results presented in Table \ref{PhotoTable} are not highly precise due to the assumed phase function and geometric albedo. Nevertheless, they are uniformly affected by these assumed values, so their relative values are highly accurate.

\begin{table*}
    \centering
    \renewcommand{\arraystretch}{1.3}
    \caption{Photometric results within apertures of fixed radii}
    \label{PhotoTable}
    \setlength{\tabcolsep}{8pt}
    \begin{tabular}{lccccccc}
        \toprule
        Date & MJD & Property$^a$ & 500 km & 1000 km & 2000 km & 4000 km & 8000 km \\
        \hline
        26 Jul 2019 & 58690 & $V$ & $21.30\pm0.03$ & $20.95\pm0.03$ & $20.51\pm0.02$ & $20.14\pm0.05$ & $19.86\pm0.04$ \\
        26 Jul 2019& 58690 & $H$ & $17.78\pm0.03$ & $17.43\pm0.03$ & $16.98\pm0.02$ & $16.61\pm0.05$& $16.33\pm0.04$ \\
        26 Jul 2019 & 58690& $C_{\rm e}$ & $1.87\pm0.06$ & $2.57\pm0.06$ & $3.88\pm0.07$ & $5.46\pm0.26$ & $7.08\pm0.29$ \\
        3 Aug 2019& 58698 & $V$ & $20.03\pm0.02$ & $19.67\pm0.01$ & $19.23\pm0.03$ & $18.86\pm0.03$ & $18.58\pm0.02$ \\
        3 Aug 2019 & 58698& $H$ & $17.50\pm0.02$ & $17.14\pm0.01$ & $16.71\pm0.03$ & $16.34\pm0.03$ & $16.06\pm0.02$ \\
        3 Aug 2019 & 58698& $C_{\rm e}$ & $2.41\pm0.05$ & $3.35 \pm 0.04$ & $5.02 \pm 0.19$ & $7.06 \pm 0.26$ & $9.14 \pm 0.26$ \\
        13 Aug 2019& 58708 & $V$ & $20.60\pm0.03$ & $20.22\pm0.02$ & $19.78\pm0.02$ & $19.41\pm0.02$ & $19.08\pm0.03$ \\
        13 Aug 2019& 58708 & $H$ & $17.27\pm0.03$ & $16.89\pm0.02$ & $16.45\pm0.02$ & $16.08 \pm0.02$& $15.75\pm0.03$ \\
        13 Aug 2019& 58708 & $C_{\rm e}$ & $2.97 \pm 0.09$ & $4.21 \pm 0.10$ & $6.31 \pm 0.10$ & $8.87 \pm 0.17$ & $11.35 \pm 0.35$ \\
        25 Aug 2019& 58720 & $V$ & $19.78\pm0.03$ & $19.38\pm0.02$ & $18.91\pm0.03$& $18.42\pm0.02$ & $18.08\pm0.03$ \\
        25 Aug 2019& 58720 & $H$ & $17.28\pm0.03$ & $16.88\pm0.02$ & $16.41\pm0.03$ & $15.93\pm0.02$ & $15.59\pm0.03$ \\
        25 Aug 2019& 58720 & $C_{\rm e}$ & $2.94 \pm 0.07$ & $4.26 \pm 0.09$ & $6.60 \pm 0.22$ & $10.22 \pm 0.22$ & $14.03 \pm 0.47$ \\
        31 Aug 2019& 58726 & $V$ & $19.26\pm0.05$ & $18.97\pm0.03$ & $18.60\pm0.02$ & $18.23\pm0.03$ & $17.95\pm0.02$ \\
        31 Aug 2019& 58726 & $H$ & $16.67\pm0.05$ & $16.39\pm0.03$ & $16.02\pm0.02$ & $15.65\pm0.03$ & $15.37\pm0.02$ \\
        31 Aug 2019 & 58726& $C_{\rm e}$ & $5.15 \pm 0.23$ & $6.69 \pm 0.21$ & $9.38 \pm 0.17$ & $13.13 \pm 0.30$ & $17.07 \pm 0.30$ \\
        \hline
        \multicolumn{8}{l}{$^a$ $V$: apparent magnitude, $H$: absolute magnitude, $C_{\rm e}$: effective scattering cross-section in km$^2$.}\\ 
    \end{tabular}
\end{table*}

\begin{figure}
\centering	\includegraphics[width=1\columnwidth]{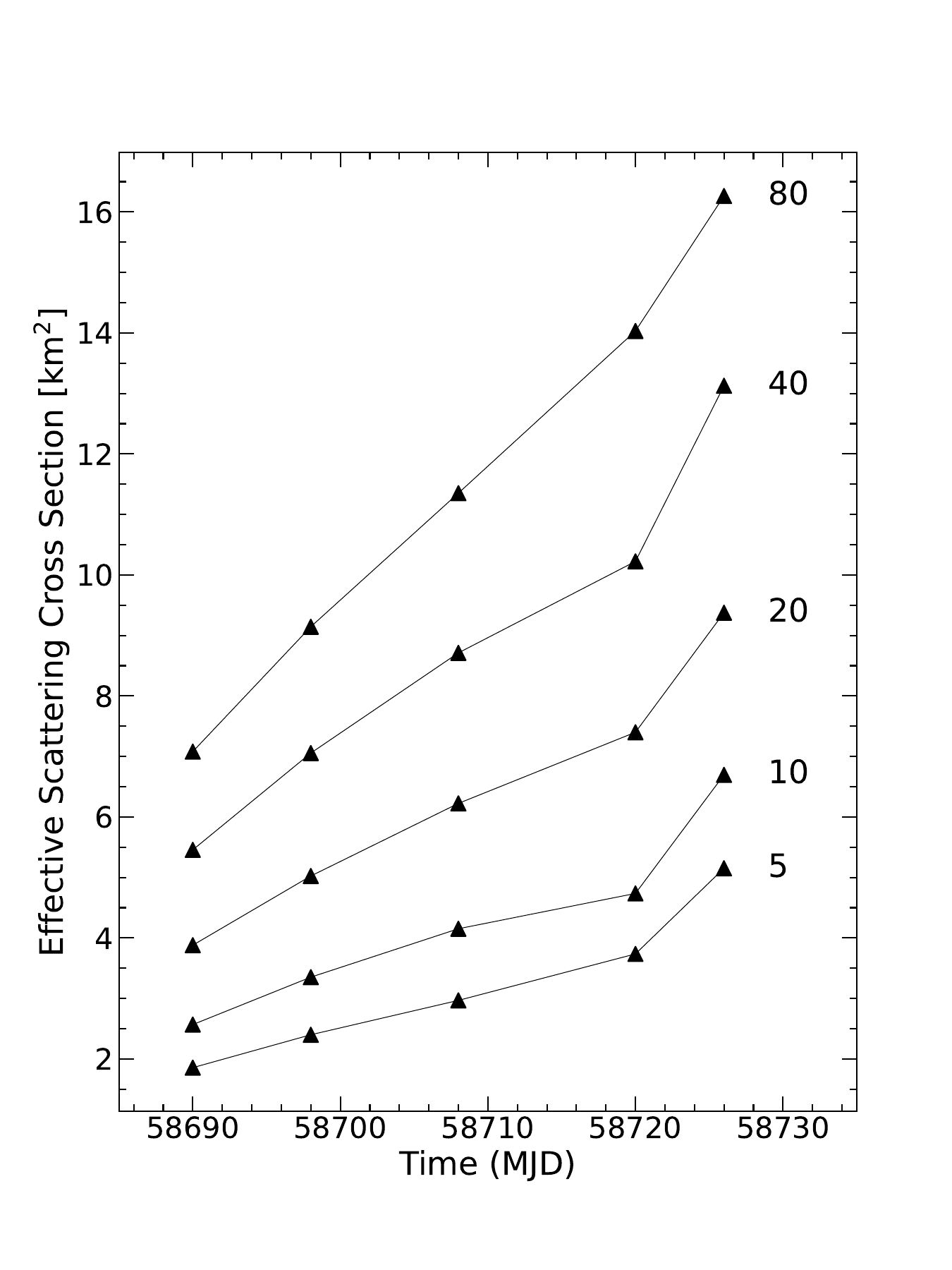}
    \caption{Effective scattering cross-section vs.~date of observation (MJD) for each of the five apertures with different radii. The radius of each aperture measured in units of $10^3$ km is presented on the right side of each line. Due to the smaller size of the error bars compared to the data points, they become concealed behind these points.}
    \label{area}
\end{figure}

The observed variations in effective scattering cross-section ($C_\text{e}$) as shown in Figure \ref{area} reflect the net effect between the dust released from the nucleus of LH$_{15}$ and the loss of previously emitted dust from the edges of the photometric aperture due to the outward expansion of the coma. An increase in $C_\text{e}$ suggests a net influx of particles, where the count of particles entering the aperture exceeds that of those exiting. Figure \ref{area} shows that $C_\text{e}$ increases nearly linearly over the observation period. This suggests that the amount of dust within the aperture is continuously accumulating, rather than reaching a steady state. Calculations based on the velocity-size relationship from Section \ref{morphology} and the parameters in Table \ref{model para} show that large dust particles remain within the aperture for the entire observation period.
These large particles with low ejection velocities, insensitive to radiation pressure, likely account for the observed increase in $C_\text{e}$. Moreover, considering small changes of LH$_{15}$'s heliocentric distance during the observation period, it is unlikely that an increase in dust production rate due to changes in the heliocentric distances causes the observed increase in scattering cross-section area.

Furthermore, Figure \ref{area} reveals that the slopes of the $C_\text{e}$ lines vary logarithmically with the radii of the apertures. This dependence arises from changes in the minimum dust particle size within the aperture. To interpret this relationship, the effective scattering cross-section is expressed as:
\begin{equation}
C_\text{e} = \int_{a_\text{min}}^{a_\text{max}} n(a) \sigma(a) \, \mathrm{d}a \, ,
\label{cro_sum}
\end{equation}
where $n(a)\mathrm{d}a=\Gamma a^{-q}\mathrm{d}a$ represents the dust size distribution, with the parameter $\Gamma$ primarily being a function of heliocentric distance, $\sigma(a)=\pi a^2$ is the scattering cross-section of a particle with the radius of $a$. The maximum radius $a_\text{max}$ is constant, and the minimum radius $a_\text{min}$ is related to the time interval $t$ required for a particle to reach the edge of the aperture with radius $r$, which is expressed as
\begin{equation}
a_\mathrm{min} = \frac{G M_\odot}{2 r R^2} t^2,
\label{cro_area}
\end{equation}
in which $G$ is the gravitational constant, $M_\odot$ is the mass of the Sun, and $R$ is the heliocentric distance. The distribution of asteroidal dust typically follows a power-law with an exponent between 3 and 3.5 (e.g.~\citet{gorkavyi2022reduction}). In this context, we assume an exponent of 3 for the dust distribution within the aperture. It can be obtained from Equation \ref{cro_sum}
\begin{equation}
\int_{a_{\text{min}}}^{a_{\text{max}}} n(a) \sigma(a) \, \mathrm{d}a = \Gamma(t)[\ln{a}]^{a_\mathrm{max}}_ {a_\mathrm{min}}.
\end{equation}
Therefore, the slopes of the $C_e$ lines is expressed as 
\begin{equation}
\frac{\mathrm{d} C_\text{e}(t, r)}{\mathrm{d}t} = \left(\ln{\frac{kr }{t^2}}\right) \frac{\mathrm{d} \Gamma(t)}{\mathrm{d}t} - \frac{\Gamma(t)}{t}, 
\label{ce_devia}
\end{equation}
where $k=\frac{2a_\mathrm{max} R^2}{G M_\odot}$ is a constant.
Thus, the changing rate of the scattering cross-sectional area with respect to time exhibits a nearly logarithmic dependence on the photometric aperture.

Figure \ref{area} shows an increase in the effective scattering cross-section of about $9.99\pm0.59$ km$^2$ over the span of about a month within the aperture of the radius of 8000 km, which suggests a rate of the effective scattering cross-section of $\frac{ \mathrm{d} C_\text{e}}{ \mathrm{d}t} = 0.28\pm0.02$ km$^2$ day$^{-1}$. It should be noted that the magnitude of the derived $\frac{ \mathrm{d} C_\text{e}}{ \mathrm{d}t}$ is influenced by the assumed phase function, which introduces uncertainty into the inferred initial time and may affect its reliability. Assuming a dust density of 1600 kg/m$^3$ and an mean radius of the particles of 0.3 mm (Table \ref{model para}), the corresponding dust production rate is derived to be about $1.56\pm0.11$ kg/s. Additionally, the rate of change of the effective scattering cross-section within apertures of 500 km, 1000 km, 2000 km, and 4000 km radii are determined as $0.09\pm0.01$ km$^2$, $0.11\pm0.01$ km$^2$, $0.16\pm0.01$ km$^2$, and $0.21\pm0.02$ km$^2$, respectively. By extrapolation of the observed trend in the effective scattering cross-section over time, it is obtained that $C_{\rm e}$ would be approximately 0 km$^2$ about $60\pm3$ days before the final observation on August 31, corresponding to July 2. This inferred initial time serves as a preliminary estimate and could be cross-validated with results from the dynamical modeling method described later.

\begin{figure*}
\centering
\begin{minipage}{0.65\textwidth} 
    \centering
    \includegraphics[width=1\columnwidth]{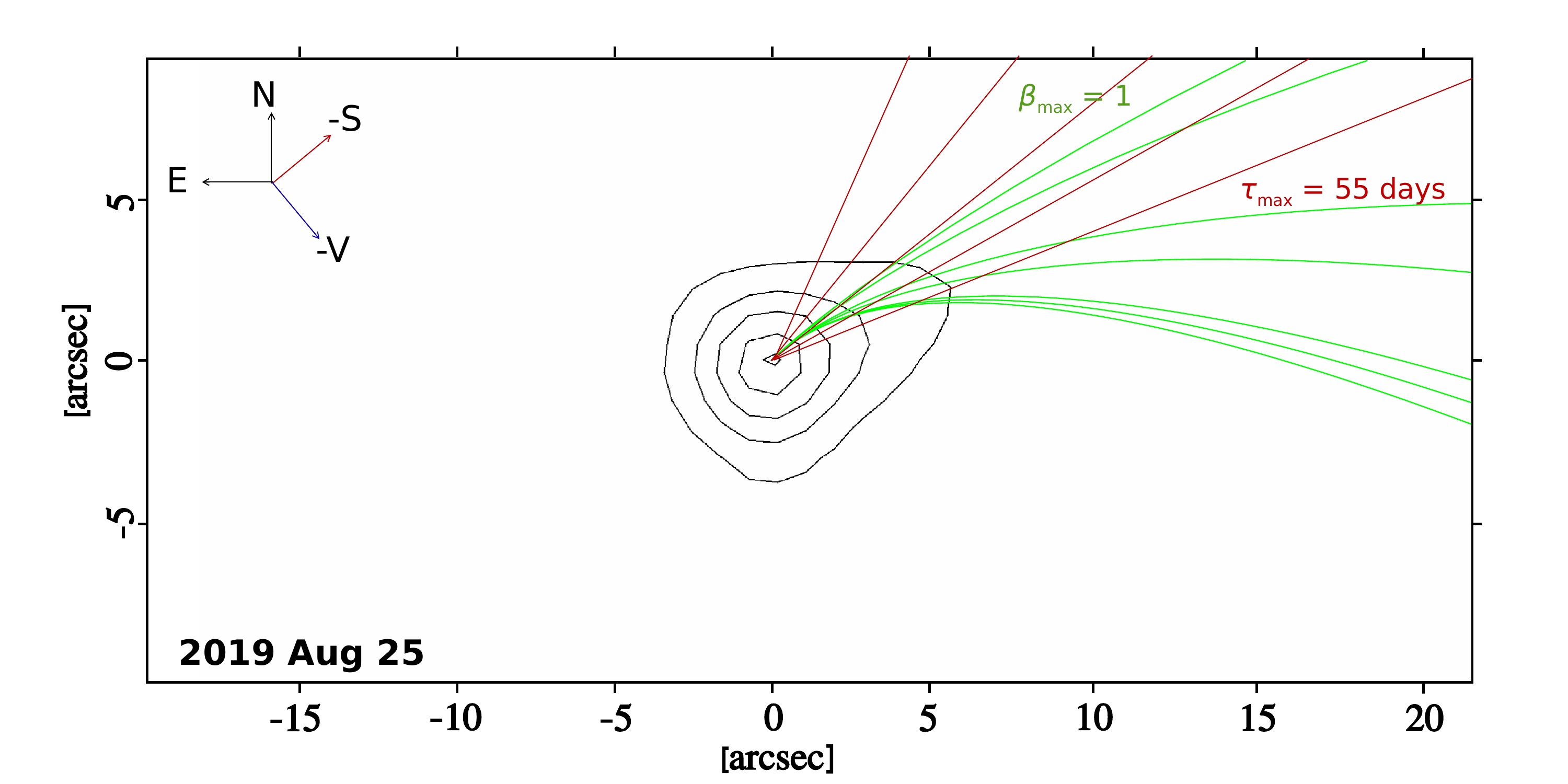}\\
    \vspace{-0.1cm}
    \includegraphics[width=0.95\columnwidth]{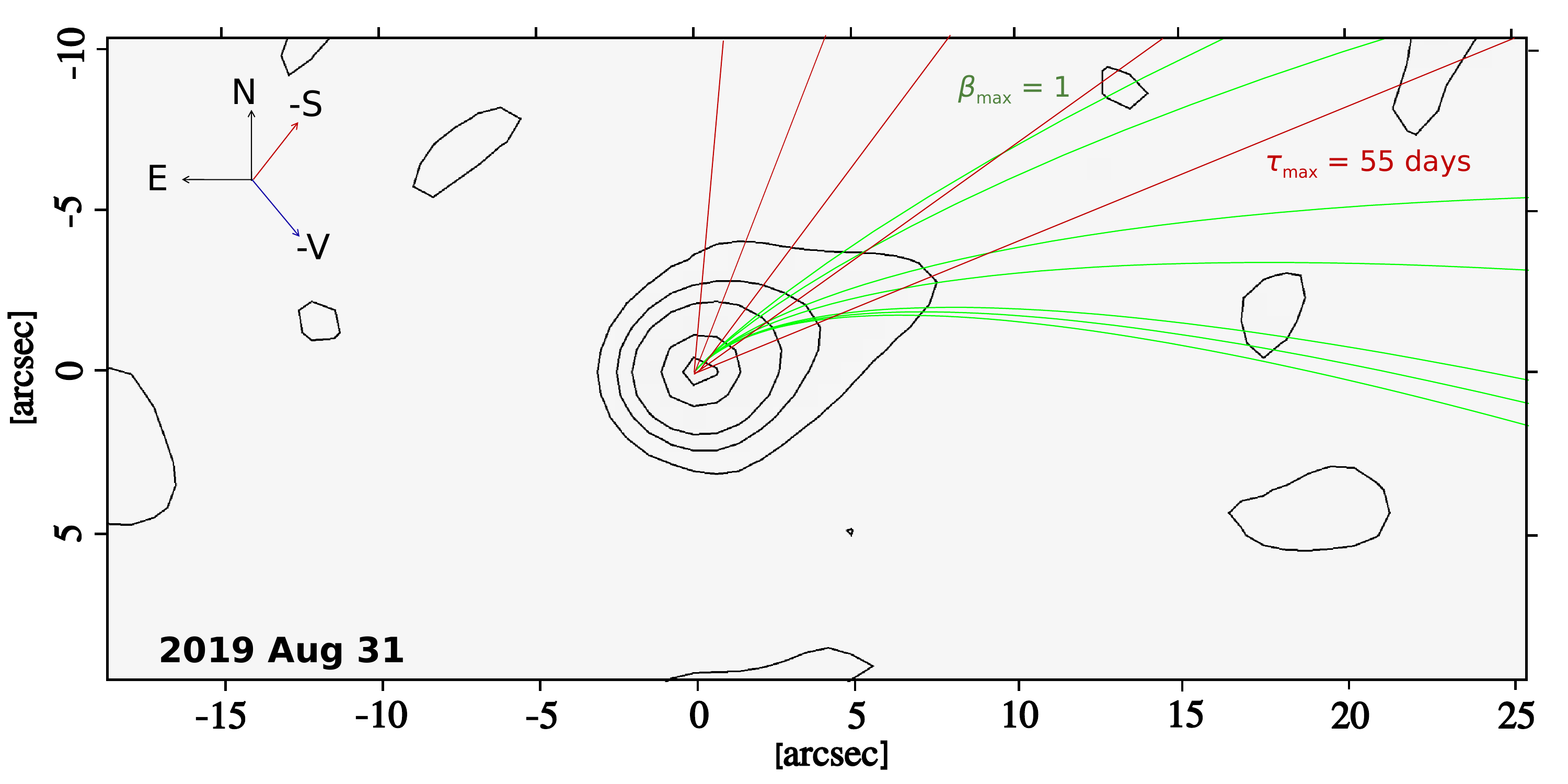}
\end{minipage}%
\begin{minipage}{0.35\textwidth} 
    \vspace{-1cm} 
    \caption{Syndynes (green) and synchrones (red) for the observation images of LH$_{15}$. The synchrones are set for dates of 55, 45, 35, 25, and 15 days prior to the observation date and the sydynes are for $\beta = 1, 1\times10^{-1}, 1\times10^{-2}$, $1\times10^{-3}$, $1\times10^{-4}$, $1\times10^{-5}$, and $1\times10^{-6}$ for LH$_{15}$ on 25 August and 31 August of 2019. The projected Sun-LH$_{15}$ radial direction (-S) and the anti-velocity direction of LH$_{15}$’s projected motion (-V) are depicted in the top left corner of each panel.} 
    \label{syncurve}
\end{minipage}
\end{figure*}
\subsection{Dust properties}
\label{Dust properties}

The physical properties of the ejected dust particles could be approximated by a morphological analysis of the dust coma and tail using the syndyne-synchrone model, which is based on the Finson-Probstein theory. The Finson-Probstein theory assumes that once dust grains are ejected from the parent body, their trajectories are influenced solely by $\beta$, which is the ratio of the solar radiation pressure to the solar gravity given by $\beta = 0.57 Q_{\mathrm{pr}}/(\rho a)$. The solar radiation pressure efficiency $Q_{\mathrm{pr}}$ is assumed as 1, and $a$ is the radius of the particle expressed in microns. In the following analysis, a density value of $\rho = 1.7\, \text{g/cm}^3$ representing the average density of 11 C-type asteroids is adopted as described by \citet{vernazza2021vlt}. A synchrone represents the collection of positions for particles of varying sizes that are ejected simultaneously, and a syndyne represents the collection of positions for particles of single sizes that are ejected continuously. For each observed epoch, the syndyne-synchrone model is computed and subsequently compared to the coma and tail in the corresponding observation image. The comparison results for all epochs show great consistency. Figure \ref{syncurve} presents the best match results overplotted on the observation images. The syndynes with $\beta$ of 0.01 and 0.0001 closely match the observed morphology of the dust tail, and the synchrones representing the end of June 2019 ($55\pm5$ days prior to the observation date) almost closely aligns with the lower boundary of the dust tail. This derived date is generally consistent with the onset date (July 2) of the activity obtained from Figure \ref{area}. The relationship between $\beta$ and radius is approximated as $a = 0.3/\beta$, indicating that the radii of the dust particles composing the dust tail and coma range between \SI{30}{\um} and 3 mm. The onset of dust activity of LH$_{15}$ occurred at a heliocentric distance of about 1.9 AU.

\subsection{Tail dynamical modeling}
\label{Tail Dynamical Modelling}

To constrain the dust environments of LH$_{15}$, the dynamical evolutions of the ejected particles are simulated by a procedure. This procedure is derived from \citet{liu2016dynamics} and has been adapted for this study. The model assumes that once a dust grain leaves the surface of a parent body it is governed by solar gravity and solar radiation pressure. Thus, if the ejection time, the initial state of the particles, and the size-dependent dimensionless factor of the dust particles are known, their positions at given observation date are uniquely determined in a heliocentric inertial system and then projected to the sky plane of the observer. The particles' contribution to the brightness of the grids in the sky plane system is computed. The spatial scale of the grid cells defined in the sky plane system is the same as that in the photographic system.

In this model, the simulated results are adjusted for the dust size limits ($\beta_\text{min}$ and $\beta_\text{max}$), the start and end times of the activity ($t_1$ and $t_0$), the ejection velocity ($v_\text{ej}$), the dust size distribution index ($s$), and the dust production rate ($\dot{M}$) to match the observational data. According to \cite{mastropietro2024photometric}, the rotation period of 2010 LH$_{15}$ is not less than 2 hours. Therefore, we assume that a "slow rotator" model for the cometary activity, which means that each point on the surface of the asteroid is in instantaneous thermal equilibrium with insolation, with the sub-solar point being the hottest. In this context, it is assumed that the dust activity is concentrated at the sub-solar point, and particles are ejected within a sunward cone with a semi-opening angle ($\omega$). To minimize computational load, preliminary constraints are applied to the input parameters before the simulation. The ejection velocity is given by $v_\text{ej} = v_0 \beta^{\gamma}$, where $v_0$ is explored around the value of $68.4\pm4.6$ m/s as detailed in section \ref{morphology}. The syndyne-synchrone analysis provides insights into the estimations of the dust size and the initial time of the activity. Figure \ref{syncurve} demonstrates that most of the particles composing the dust tail align with syndynes characterized value of $\beta$ of less than 0.01, and the start time of the activity ($t_1$) is most closely approximated by the synchrone corresponding to end of June 2019. The continuous increase in the effective scattering cross-section in Figure \ref{area} implies that the dust activity has been ongoing; therefore, the end time $t_0$ is set to coincide with the epoch of observation. It is assumed that the size distribution index $s$ is 3.5. The remaining parameters $\beta_\text{min}$, $\omega$, and the dust production rate, $\dot{M}$, are treated as free variables within this model. To ensure an accurate comparison with the observed image, the modeled images are convolved with a Gaussian function whose FWHM is adjusted to match the seeing conditions at the observation epoch.

\begin{figure}
\centering
	\includegraphics[width=1\columnwidth]{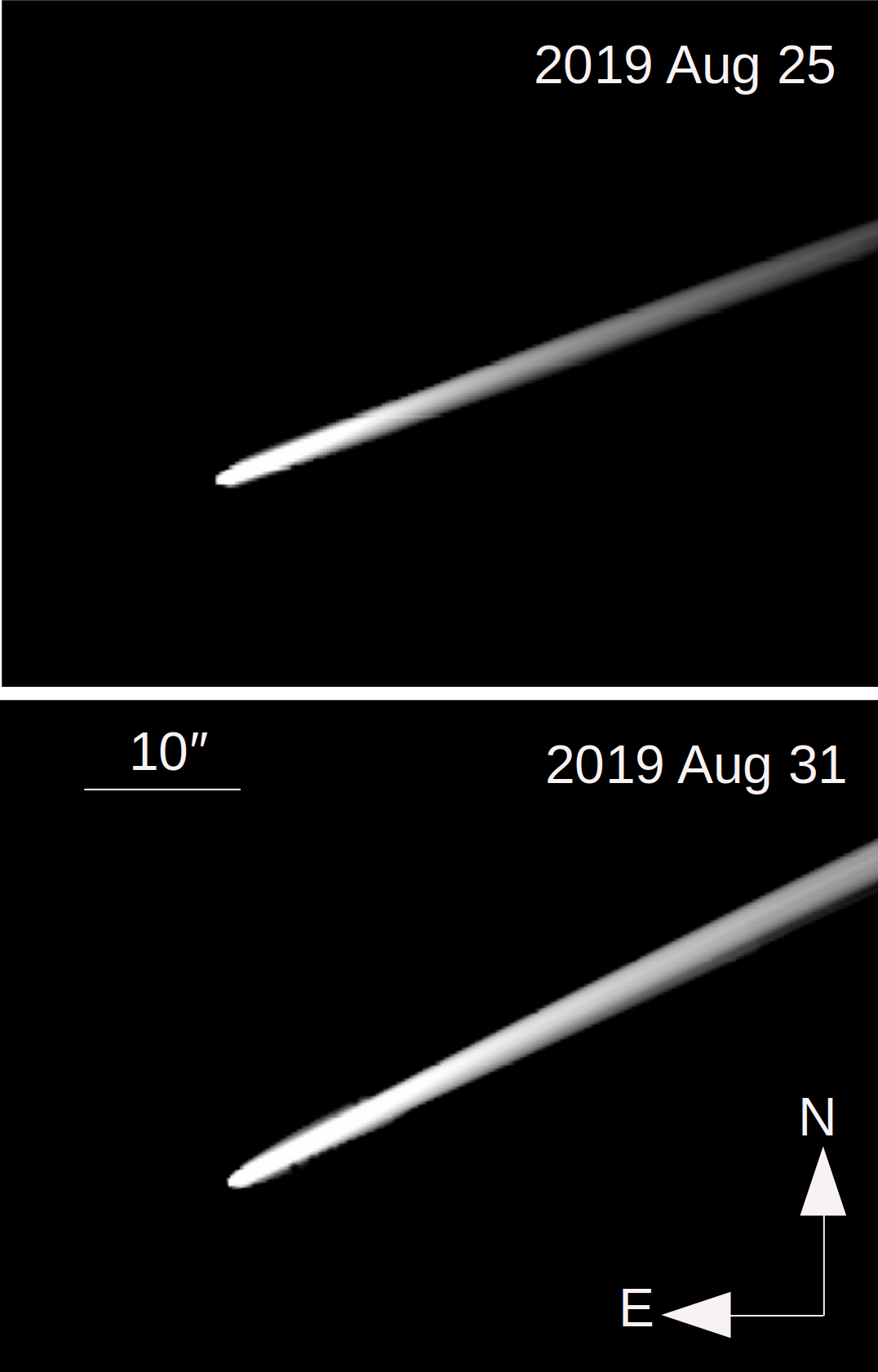}
    \caption{The modeled morphology of LH$_{15}$ using the parameters in Table \ref{model para}. The observation date is marked in the upper right corner of each panel. The true north direction (N) and the true east direction (E) are indicated as white arrows. The scale bar is also marked. }
    \label{dustmodel}
\end{figure}

\begin{figure*}
\centering
	\includegraphics[width=2.1\columnwidth]{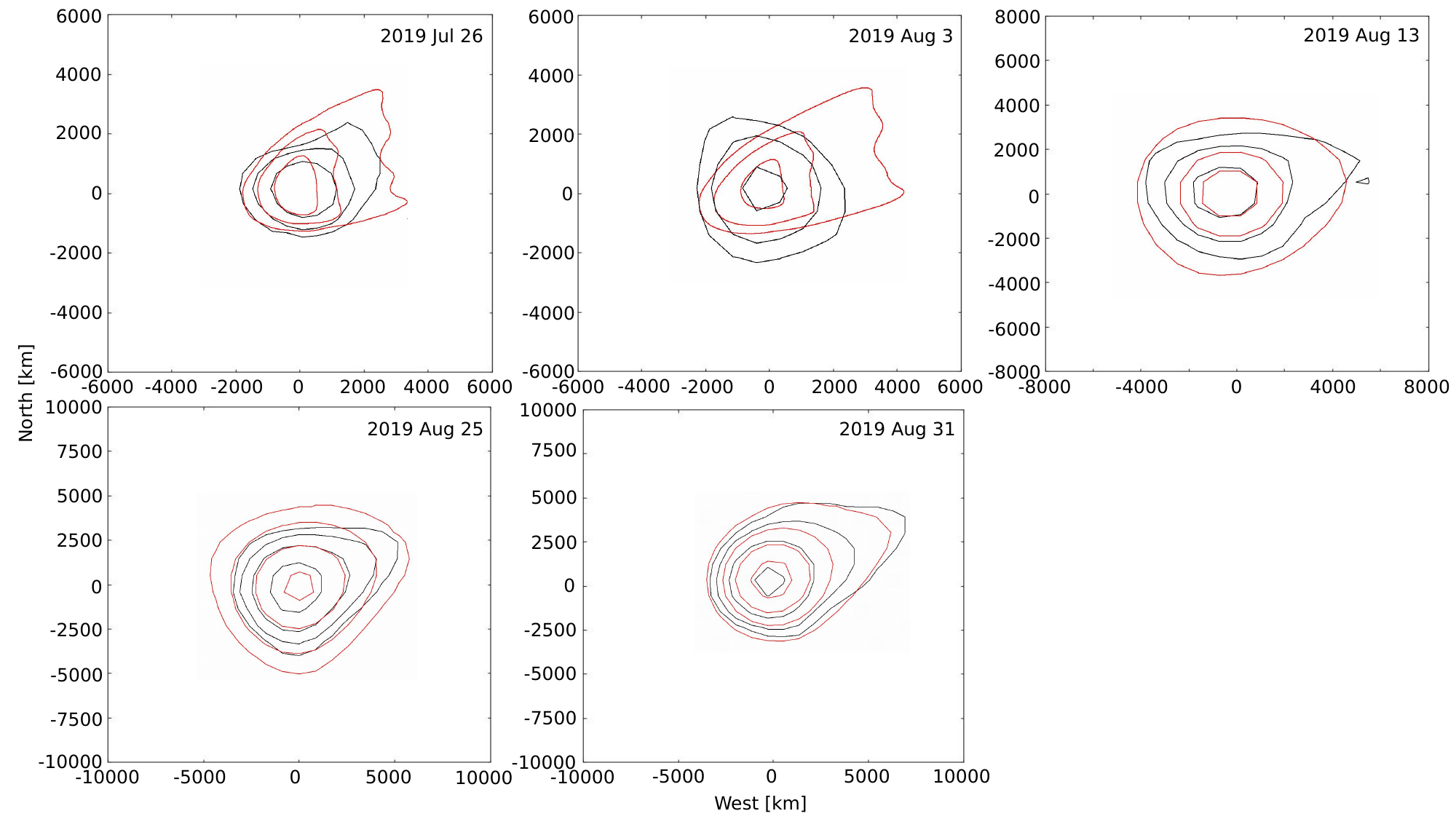}
    \caption{Comparison between brightness contours in observed images (black lines) and modeled images (red lines, using the parameters in Table \ref{model para}) from 26 July to 31 August, 2019. The brightness levels corresponding to the innermost contours in the images for each of the five dates are 23.83 mag arcsec$^{-2}$, 23.11 mag arcsec$^{-2}$, 22.99 mag arcsec$^{-2}$, 22.39 mag arcsec$^{-2}$, and 22.29 mag arcsec$^{-2}$, respectively. For adjacent contours, the brightness level of the outer brightness contour is half that of the inner brightness contour. The same step (logarithmically) is adopted for both observed and modeled images.}
    \label{Isophotes}
\end{figure*}

A series of model images are generated through simulations using a range of parameter combinations. These images are subsequently compared to the observational data to identify the optimal parameter combination. In this paper, a grid search method is chosen to find the optimal parameters for the sake of computational simplicity. For each observation epoch, the quality of the match between the observed and simulated images depends on the parameter of $\chi = \sqrt{\frac{\sum (i_{\text{obs}} - i_{\text{sim}})^2}{N}}$, where the $i_\text{obs}$ is the pixel intensity in the observation image and $i_\text{sim}$ is the pixel intensity in the model image. It should be noted that the region affected by cosmic rays and the field stars are not considered in the analysis of the pixel intensity. The model parameters are listed in Table \ref{model para}. The simulated images that best match the observations are displayed in Figure \ref{dustmodel}. Isophotes measured and modeled near the nucleus of LH$_{15}$ are presented in Figures \ref{Isophotes}.

\begin{table*}
    \centering
    \renewcommand{\arraystretch}{1.2}
	    \caption{Model parameters}
	    \label{model para}
	 \begin{tabular}{ccccc} 
		\hline
		 Parameter & Simulated Values & Best Parameter Values [Possible Range] \\
		\hline
		$\beta_\text{max}$&$10^{-2}$&fixed\\
		$\beta_\text{min}$&$10^{-6}, 10^{-5}, 10^{-4}$&$10^{-4}$ [$10^{-5}$, $10^{-4}$]\\
		$t_0$&26 May to 26 July, with an interval of 10 day& 26 June [16 June, 26 June]\\
		$t_1$&The Observation Epoch&fixed\\
		$v_0$ (m/s)&30 to 110, with an interval of 20&50 [30, 70]\\
            $\gamma$&0.1 to 0.9, with an interval of 0.2&0.5 [0.3, 0.7]\\
		$s$&3.5&fixed\\
		$\omega$ ($^\circ$)&15 to 60, with an interval of 5 & 40 [45, 35]\\
		$\dot{M}$ (kg/s)&1 to 21, with an interval of 5&6 [1, 11]\\
        \hline 
    \end{tabular}
\end{table*}

The comparison results in Table \ref{model para} show that the ejection velocity ($v = v_0\, \beta^{0.5}$, $v_0=50\pm20$ m/s) is moderately correlated with the dust size of the particles ($\gamma = 0.5\pm0.2$), exhibiting a corresponding range of from about 0.5 m/s to 5 m/s for grain radii $a$ ranging from 0.03 mm to 3 mm with $a = 0.3/\beta$. These results are generally consistent with the range of from 0.68 m/s to 6.8 m/s for grain radii ranging from 0.03 mm to 3 mm, as inferred from the morphological measurement in Section \ref{morphology}. The derived onset date of the activity is about 26 June 2019 with the heliocentric distance of about 1.86 AU, and the activity continued until the last observation epoch. The average dust mass production rate matching the observed brightness is $6\pm5$ kg/s, which is compatible with the result obtained from photometric measurements in Section \ref{photometry}. The production rate mentioned here refers to the average rate. The actual production rate is likely to be significantly higher than the derived production rate, owing to the presence of boulders that are not optically dominant. An effort is also made to reproduce the observed image using an isotropic ejection model ($\omega=180^\circ$). Nevertheless, the isotropic ejection model fail to match well to the data, as it leads to a tail with a greater width than observed. A possible scenario is that due to solar heating, sublimation on the asteroid's surface triggers the dust activity. Given that the temperature is highest near the sub-solar point, the dust grains may be preferentially ejected anisotropically within a solid cone-shaped jet from the region around this point.

The parameters of the dust activity derived from the complicated modeling presented in Table \ref{model para} are generally consistent with those obtained from the basic measurements in Sections \ref{morphology} and \ref{photometry}. This consistency across relatively independent methods enhances the robustness and reliability of our results. These findings, including the onset date, the dust production rate, and the ejection velocity, provide support for the subsequent determination of the activity-driving mechanism.

\section{Discussion}
\label{discussion}
\subsection{Activity-driving mechanism}
The photometric results in Figure $\ref{area}$, the comparison results of the dust tail in Table \ref{model para}, and the recurrent activity from 2010 LH$_{15}$ mentioned in \citet{chandler2021recurrent} indicate that the activity observed is inconsistent with an origin of a single-pulse ejection event, such as a collision. In other words, the activity of LH$_{15}$ is likely driven by alternative mechanisms capable of causing a sustained process, including rotational instability, rubbing of the binary asteroid, and sublimation.

The surface disruption of the asteroid due to the rotational instability occurs under conditions where the combined gravitational and cohesive forces acting on the grains are less than the centrifugal force. Rotational instability may also trigger multiple mass loss events like those found in the recurrent activity of LH$_{15}$, the appearance of twin-tails of (6478) Gault is a prime example \citep{kleyna2019sporadic}. However, the velocity of the particles shed from the surface of an asteroid due to rotational instability is usually near zero and has a weak size dependence, which are incompatible with the model results in Table \ref{model para} ($\gamma\approx0.5$). Current results suggest that rotational instability is less likely the origin of the activity, but it cannot be completely ruled out. Whether rotational instability serves as the cause of the activity depends on the determination of the rotation period of LH$_{15}$.

Repeated contacts between components of a binary system can also lead to multiple mass-loss events and are considered as a possible cause for the multiple tails of 311P \citep{hainaut2014continued}. However, all ejection events of 311P occurred within the span of 1 year \citep{jewitt2015episodic, liu2023active}, whereas the interval between the two ejection events of LH$_{15}$ is close to 10 years. The disparity in time scales reduces the likelihood that contact binary disruption serves as the mechanism of the activity of LH$_{15}$.

The continuity of the activity (Figure $\ref{area}$), the dependence of the ejection velocity on dust size ($\gamma\approx0.5$, Table \ref{model para}), and the recurrence of the activity (\citep{chandler2023new}) indicate that the hypothesis of the sublimation being the possible origin of the activity of LH$_{15}$ is worth further discussion. The mass flux of the sublimation of exposed ice, $f_\text{s}(T)$, is calculated using the energy balance equation from
\begin{equation}
\frac{S_{\odot}}{R^{2}}(1-A)=\chi\left[\epsilon \sigma T\left(R\right)^{4}+H_\text{s} f_\text{s}(T)\right].
	\label{waterRate}
\end{equation}

Here, the solar constant is denoted as $S_{\odot}$ and has a value of 1361 W m$^{-2}$. The heliocentric distance, $R$, is referenced in Table \ref{geometry}. The bond albedo is represented by $A$, with an assumed value of 0.04. The term $\chi$ indicates the distribution coefficients for the incident heat and is assumed as 2. The effective emissivity is set to $\epsilon = 0.9$. The Stefan–Boltzmann constant is denoted as $\sigma$ and has a value of $5.67\times10^{-8}$ W m$^{-2}$ K$^{-4}$. The variable $H_\text{s} (T)$ represents the latent heat at a particular temperature $T (R)$ at the heliocentric distance of $R$ (Table \ref{geometry}).

At a heliocentric distance of 1.77 AU on 31 August 2019, the equilibrium temperature is estimated to be about 207 K, with the mass flux of the sublimation of exposed ice $f_\text{s}$ determined from Equation (\ref{waterRate}) being $4.4\times10^{-5}$ kg m$^{-2}$ s$^{-1}$. The area that could maintain the observed dust activity ($\dot{M}=6\pm5$ kg/s; Table \ref{model para}) is estimated by

\begin{equation}
A_{\mathrm{s}} = \frac{\dot{M}}{f_\text{d2g}f_{\mathrm{s}}}.
	\label{MasswaterRate}
\end{equation}
where the mass production rate ratio of dust to water $f_\text{d2g}$ is assumed as 10 \citep{fulle2016evolution, reach2000formation, kim2022sublimation}. From Equation (\ref{MasswaterRate}), the active area is $A_{\mathrm{s}} = (1.3\pm1.08)\times10^{4}$ m$^2$, which is equivalent to 0.1\% of the total surface area of a sphere with a radius of $1.11\pm0.02$ km (Equation (\ref{radius})). Assuming the active area is circular, its radius is determined to be approximately $64\pm27$ meters.

\begin{figure}
\centering
	\includegraphics[width=1\columnwidth]{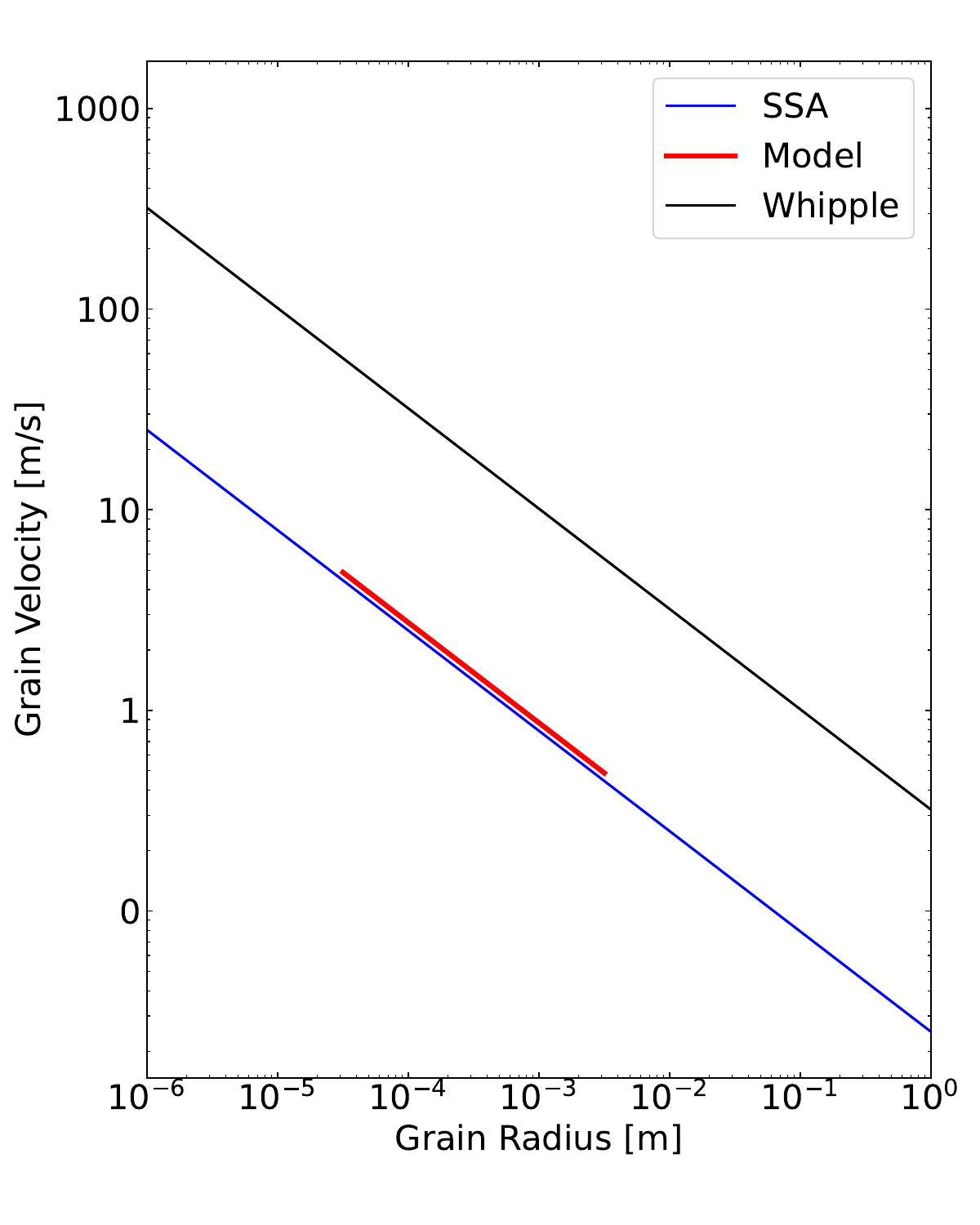}
    \caption{Comparison between the size-velocity relationship determined by this study (red line), and those determined by the classical Whipple model (black line) and the Small Source Approximation (SSA) model (blue line). The gravity of the nucleus is not considered in the models.}
    \label{velocity}
\end{figure}

In the absence of nucleus gravity consideration, the comparison between the size-velocity empirical formula determined by tail dynamical modeling in Section \ref{Tail Dynamical Modelling} with those derived from the classical Whipple model \citep{whipple1951comet} and the Small Source Approximation (SSA) model \citep{jewitt2014hubble} is shown in Figure \ref{velocity}. The difference between the classical Whipple model and the SSA model lies in the size of the active area. The Whipple model assumes that the active area encompasses the sunward hemisphere, whereas the SSA model assumes that the active area constitutes only a small fraction of the total surface area. This difference in the active area results in variations in the distance for the particle accelerated by gas drag. It is shown in Figure \ref{velocity} that the ejection velocity of the particles determined in this study is an order of magnitude lower than that determined by the Whipple model, but is comparable to that determined by the SSA model. Therefore, it is a reasonable hypothesis that the sublimation of water ice in a small active area caused the observed dust activity of LH$_{15}$.

The hypothesis of the sublimation as the mechanism of activity for LH$_{15}$ provides new information about the distribution of water ice in the main asteroid belt. The results of the dust tail dynamics modeling show that the heliocentric distance at the onset of the activity of LH$_{15}$ (1.86 AU) is smaller than that of the main-belt comet 259P (1.96 AU), which previously had the smallest heliocentric distance at the onset of the activity among known main-belt comets \citep{hsieh2021reactivation, kim2022hubble}. If the orbit of LH$_{15}$ is stable, the sublimation-driven activity of LH$_{15}$ indicates the presence of water ice at a heliocentric distance of 1.86 AU within the main asteroid belt region, resulting in an extension of the water-ice-bearing region in the main asteroid belt inward by 0.1 AU. 

\subsection{Comparison with other active asteroids}
\label{Comparison with Other MBCs}
A detailed summary of the activity parameters of the active asteroids (including 238P, 259P, 288P, 324P, 358P) suspected to be driven by the sublimation is listed in Table (3) in \citet{hsieh2021reactivation}. Disregarding measurement errors, the onset of the activity of LH$_{15}$ occurs at a heliocentric distance of 1.86 AU, which is less than $r_{h, 0}$ (heliocentric distance at the onset of the activity of other active asteroids) listed in Table 3 in \citet{hsieh2021reactivation}. However, once LH$_{15}$ is activated, its dust production rate does not differ significantly from that of other active asteroids (Table (3) in \citet{hsieh2021reactivation}). The onset of activity in LH$_{15}$ at a smaller heliocentric distance compared to those of other active asteroids indicates that the reservoir of volatiles may be buried deeper below the surface than those in other active asteroids. The longer period of energy transmission from solar radiation to the reservoir of volatiles in LH$_{15}$, compared to other main-belt comets, may result in the observed delay in onset of activity of LH$_{15}$. 

The depth at which the reservoir of volatiles is buried may also be related to the size of the asteroid. The radius of LH$_{15}$ measured in this study is observed to be more than twice that of most of active asteroids listed in Table 3 in \citet{hsieh2021reactivation}, indicating that LH$_{15}$ is likely an older asteroid. This is inferred from the fact that the larger objects have longer collisional destruction timescales, leading to the assumption that the larger objects are generally older \citep{cheng2004collisional, bottke2005linking}. Over time, the depletion of near-surface layers of volatiles may result in the reservoir of volatiles in the older objects, such as LH$_{15}$, appearing to be buried deeper compared to those in the younger objects.

\section{Conclusions}
\label{CONCLUSIONS}
The observations of the newly active asteroid 2010 LH$_{15}$ from the archive are presented and analyzed using relatively independent methods, which provide comparable results. The main conclusions are listed as follows:

1. The newly active asteroid 2010 LH$_{15}$ was discovered to exhibit the reactivation in July 2019, with the length of the dust tail measured to increase from $2.4\pm0.3$ arcseconds on 26 July to $18.5\pm1.7$ arcseconds on 31 August.

2. The observations during the inactive periods show that the absolute magnitude of the nucleus is $H_r = 17.01\pm0.04$, corresponding to a radius of $r_e=1.11\pm0.02$ km with assumed geometric albedo of $p_r = 0.05$. The color index of the nucleus is relatively close to that of the ejecta around the nucleus, with a value of $H_g - H_r = 0.44\pm0.07$.

3. The results of dust photometry indicate that the activity was consistently ongoing throughout the observation period, and the effective scattering cross-section increases at an average rate of $0.28\pm0.02$ km$^2$ day$^{-1}$.

4. The best-match parameter set for the simulated images with observed images indicates that dust activity might start as early as 26 June 2019, with the ejected dust particles having a radius ranging from 0.03 mm to 3 mm, and a corresponding velocity range of 0.5 m/s to 5 m/s.

5. The recurrence of the activity near perihelion, the continuity of the activity, and the dependence of the ejection velocity on dust size ($\gamma\approx0.5$) collectively support the hypothesis of the sublimation being the possible origin of the activity. 

6. The average dust mass production rate of the activity is $6\pm5$ kg/s, which is supported by an equilibrium sublimation area of about $A_{\mathrm{s}} = (1.3\pm1.08)\times10^{4}$ m$^2$. 

7. The heliocentric distance at the onset of the activity of LH$_{15}$ is less than that of other active asteroids which are suspected to be driven by the sublimation, indicating that the reservoir of volatiles of LH$_{15}$ is buried deeper than that of these active asteroids.

\begin{acknowledgements}
This work was supported by the National Natural Science Foundation of China (No.~12472048, 12311530055, and 12002397), and the National Defense Science and Engineering Bureau civil spaceflight advanced research project (No.~D050201). The data used in this work is generated as detailed in the text and will be shared on reasonable request to the corresponding author.
\end{acknowledgements}

\bibliography{aanda}{}
\bibliographystyle{aa}

\end{document}